\documentclass[11pt]{article}

\usepackage{jheppub} 
                     
\usepackage{braket,slashed,bm}

\usepackage{array,multirow}
\usepackage[normalem]{ulem}
\usepackage[T1]{fontenc} 
\usepackage{soul}

\usepackage{mathrsfs}

\usepackage{tikz}
\usetikzlibrary{arrows,decorations.pathmorphing,backgrounds,positioning,fit,petri,automata,shadows,calendar,mindmap,
decorations.markings,calc}

\newcommand{\be}{\begin{equation}}
\newcommand{\ee}{\end{equation}}
\newcommand{\bea}{\begin{eqnarray}}
\newcommand{\eea}{\end{eqnarray}}
\newcommand{\ba}{\begin{array}}
\newcommand{\ea}{\end{array}}

\newcommand{\nn}{\nonumber}

\title{Effective interpretations of a {diphoton} excess}

\author[a]{Laure Berthier,}

\author[a,b]{James M.\ Cline,}

\author[a]{William Shepherd,}

\author[a]{Michael Trott}

\affiliation[a]{Niels Bohr International Academy \& Discovery Center, 
Niels Bohr Institute, University of Copenhagen, 
Blegdamsvej 17, DK-2100, Copenhagen, Denmark}
\affiliation[b]{Department of Physics, McGill University,
3600 Rue University, Montr\'eal, Qu\'ebec, Canada H3A 2T8}

\abstract{We discuss some consistency tests that must be passed for a
successful explanation of a {diphoton} excess at larger
mass scales, generated by a scalar or {pseudoscalar}
state, possibly of a composite nature, decaying to two photons. 
{Scalar states at mass scales above the electroweak scale decaying significantly into photon final states generically
lead to modifications of Standard Model Higgs phenomenology. We characterise this effect using
the formalism of Effective Field Theory (EFT) and study the modification of the effective couplings to photons and gluons of the Higgs.
The modification of Higgs phenomenology comes about in a variety of ways.
For scalar $0^+$ states, a component of the Higgs and the heavy boson can mix.  
Lower energy phenomenology gives a limit on the mixing angle, which gets
generated at one loop in any theory explaining the diphoton 
excess.  Even if the mixing angle is set to zero, we demonstrate that a relation exists between lower energy Higgs data 
and a massive scalar decaying to diphoton final states.
If the new boson is a pseudoscalar, we note that if it
is composite, it is generic to have an excited scalar partner that 
can mix with a component of the Higgs, which has a stronger coupling
to photons. In the case of a pseudoscalar, we also characterize how 
lower energy Higgs phenomenology is directly modified using EFT, even without 
assuming a scalar partner of the pseudoscalar state. We find that naturalness concerns can be accommodated, and that
pseudoscalar models are more protected from lower energy constraints.}}

\begin{document} 
\maketitle

\section{Introduction}\label{sec:intro}

The global data set reported by LEP, the Tevatron, LHC and a host of
low-energy experiments is consistent with the Standard Model (SM) of
particle physics. With the discovery of a $0^+$ scalar ($h$)
consistent in its properties with the scalar $0^+$ component of the SM Higgs doublet ($H$), {any} 
extension of
the SM that aims to explain new phenomena {is constrained by}
an even larger bevy of lower energy tests. 
With the initial reporting of {run I}I data at $\sqrt{s} \cong 13 \, {\rm TeV}$, lower energy tests now include the
properties of the "Higgs pole" measurements, fixed to $m_h \cong 125 \, {\rm GeV}$, measured at $\sqrt{s} \cong 7,8 \, {\rm TeV}$ in 
{run I}. In this paper, we discuss a set of consistency
conditions for scalars with mass scales $m_s \gg m_h$ that
generate a significant decay to diphoton final states, {arising from} these lower energy measurements.
We will assume that the 125 {\rm GeV} scalar is 
{approximately} the SM Higgs boson and study the perturbation of its properties
using the Standard Model Effective Field Theory (SMEFT) formalism. 
The modification of the SM Higgs properties comes about in a variety of ways.
For example, a component of the SM Higgs can mix with a new $0^+$ scalar or higher-mass resonances.  
The constraints we derive on the mixing from the
experimentally established Higgs couplings must be respected by
any models with new scalars that can mix significantly
with the Higgs.  Other constraints, not directly tied to mixing, are also present
when studying low-energy phenomenology. We characterize these matching effects using the SMEFT.

Our motivation is the report of a slight excess of {diphoton}
events in the ATLAS and CMS data
\cite{ATLAS-CONF-2015-081,CMS-PAS-EXO-15-004} at $\sim 750$ {\rm GeV}.
This excess {might} be, and arguably most likely
{is}, a statistical
fluctuation \cite{ATLAS-CONF-2015-081,CMS-PAS-EXO-15-004,LEE}.\footnote{We note that the
arguments we advance are quite general for higher scale composite
resonances that have a significant branching fraction to {diphoton}
final states, even if the current excess is a statistical fluctuation.}
However, the possibility that this excess is generated by new physics
has received a lot of attention. {Many authors have considered models in which the hypothesized
scalar is composite, due to the need for it to couple to gluons or
photons despite it being neutral.  It is interesting to
consider what effects such a state could have on the observed
properties of the Higgs boson.}  Mixing 
{of} {{$h$} {with} any new states that decay to {diphoton}s
will
introduce a shift in the expected branching ratio 
{for $h\to\gamma\gamma$}. With the
measurements of {run I}, it is known that any such
{perturbation cannot greatly alter the observed branching
ratio, which is $B(h\to\gamma\gamma)\cong 2\times 10^{-3}$.}
{Numerous} higher dimensional
operators at lower scales have also been probed at LHC in {run I},
and in Electroweak Precision Data (EWPD) studies. We study the
consequences of  a {diphoton} excess at $\sim 750$
{\rm GeV} in a wide class of models {arising from} consistency with these
lower energy tests.

The purpose of this paper is to further develop these consistency
tests and to apply them to {generic} 
models that could 
explain {the putative} excess. 
{Although some of the constraints we will derive can be satisfied by
choosing parameters such that the scalar-h mixing angle is
sufficiently
small in some models, it is interesting to ask whether such values
are natural or if they require fine tuning.  This issue is sharpened
by the fact that} 
the mixing of interest is {necessarily} generated by the same operators
that are assumed to exist {for
the purpose of explaining the diphoton events}. Other (weaker) constraints we derive are not related to the 
scalar-h mixing angle at all, but still must be respected.

{On the other hand,
{pseudoscalar} states are forbidden by parity from mixing with $h$.}  
 However, we will argue that {pseudoscalar} states
{in the spectrum} of a strongly confining sector are likely to be
accompanied by scalar states, with an even stronger (effective) coupling to photons,
on fairly general grounds, { leading to indirect constraints on
sectors with composite pseudoscalars as the lightest states.} Further, we characterize how
pseudoscalar states still lead to modified properties of the SM Higgs in lower energy experiments using the formalism
of the SMEFT.
{The conditions we develop provide a}
challenge to {the construction of}
consistent strongly interacting
models {for the diphoton} excess.

\subsection{Properties of the diphoton excess at $\sim 750 {\rm GeV}$} The properties of the $\sim750$ $\rm GeV$ diphoton excess have been reported in detail by the experimental
collaborations \cite{ATLAS-CONF-2015-081,CMS-PAS-EXO-15-004}. In brief
summary, the excess at  $\sim 750$ $\rm GeV$ in diphoton
final states is characterised as resonant production with an
approximate cross section $\sigma(pp \rightarrow S \rightarrow \gamma
\, \gamma)\approx8[fb]$. The excess in ATLAS data has 
a local statistical significance of 
$3.6\,\sigma$ and global significance of $2.0\,\sigma$, while that of
CMS is at 
$2.6\,\sigma$, with a global significance of only $1.2\,\sigma$.
The two experiments have differing
preferences for the width of the resonance, with CMS
preferring a narrow {state} {relative} to the experimental resolution
of about 6 GeV, while
the ATLAS data {prefer} a larger width of around $45 \, {\rm GeV}$.
For this reason we will formulate our consistency conditions in a manner that allows
the width to be easily adjusted, to a future experimental value, that is more
consistent between the experimental results.
\section{Scalar models}
The Landau-Yang theorem \cite{Landau:1948kw,Yang:1950rg}
{states} that a resonance decaying to
diphotons {can only have spin 0 or
spin 2}. {Here} we do not consider spin 2 models, or the simultaneous production of other,
undetected, states to consider other possibilities. Spin zero 
particles can be either scalar or {pseudoscalar},
and either fundamental or composite. We first focus on the scalar case.

In a fairly general class of models,
the scalar field $S$  couples to gluons, photons and 
{possibly} quarks in order to {explain the production and 
decay of $S$ into}  photons {that give the}
 diphoton excess. 
For a scalar {of} mass $m_s$  and width $\Gamma_s$,
one can {express} the extra (due to $S$) contribution to the cross
section times branching ratio {to photons as}
\bea
\label{delta_sigma}
\Delta \sigma(pp \rightarrow S \rightarrow \gamma \, \gamma) &=&
 \frac{\Gamma (S \rightarrow \gamma \, \gamma)}{m_s \, \Gamma_s \, s}
 \left[\mathcal{C}_{gg} \, \Gamma(S \rightarrow gg) +
 \mathcal{C}_{\gamma \, \gamma} \, 
\Gamma(S \rightarrow \gamma \,\gamma) + \sum_q \, \mathcal{C}_q  
\, \Gamma (S \rightarrow q \, \bar{q}) \right]. \nn\\
&=& 2.3{\rm\,pb} \times 
 \frac{\Gamma (S \rightarrow \gamma \, \gamma)}{m_s}
	\sum_i {\cal C}_{i}\, {\rm Br} (S\to i\,i)
\eea
For $\sqrt{s} = 13 \, {\rm TeV}$ the dimensionless {coefficients are approximately} 
\bea
\mathcal{C}_{i} = \{\mathcal{C}_{\gamma\gamma}, \mathcal{C}_{b}, \mathcal{C}_{c},\mathcal{C}_{s}, \mathcal{C}_{u}, \mathcal{C}_{d}, \mathcal{C}_{gg}\} \simeq \{0.53, 15.3, 35.7, 83, 1054, 627, 2137 \}.
\eea
For example if
 $\Gamma(s \rightarrow \gamma \, \gamma) \sim 
\Gamma(s \rightarrow g g) \sim 0.01\,m_s$, {we find}
$\sigma\sim [fb] \times \mathcal{C}_i$.
The {$\cal{C}_{\gamma\gamma}$} term 
{was} recently derived in \cite{Csaki:2015vek} using the 
equivalent photon approximation, assuming that the inverse of the 
impact parameter scaled to
the proton radius is $r_\star \sim 0.13$.\footnote{This result is similar to the elastic scattering result reported in Ref.~\cite{Fichet:2015vvy}, which also reports inelastic
scattering results, which are dominant. See the Appendix for further discussion on this point.} The {latter coefficients} 
were generated in 
{Ref.}\ \cite{Franceschini:2015kwy} {at}
 a renormalization scale $\mu = m_s$ using MSTW2008 
{parton distribution functions (PDFs)}
 \cite{Martin:2009iq}. The parton luminosities are such 
that gluonic or photonic production of the state can dominate.
Utilizing the quark production 
mechanism has been examined in Ref.\ \cite{Aloni:2015mxa}, and found to
be challenging.

We focus on the cases of production and decay through 
$gg \rightarrow S$ and $S \rightarrow \gamma \, \gamma$.
{We consider}  the case 
{where} the scalar field $S$ {couples
via} the operators\footnote{Of course in a general scalar singlet case, all dimension five operators of the form $S \, \times \mathcal{L}_{SM}$ are present. And considering dimension six operators
many other operators, for example, $S^2 B^{\mu \, \mu} \, B_{\mu \, \nu}$ are also present. Our purpose is to link a high energy diphoton excess in a minimal scenario with lower energy phenomenology, so these further Lagrangian terms with unknown Wilson coefficients are neglected.}
\bea\label{basicL}
\mathcal{L}_{int} = \frac{c_G \, g_3^2}{\Lambda_{g}} S \, G^{\mu \, \nu} \, G_{\mu \, \nu} + \frac{c_B \, g_1^2}{\Lambda_{\gamma}} S \, B^{\mu \, \nu} \, B_{\mu \, \nu} +  \frac{c_W \, g_2^2}{\Lambda_{\gamma,2}} S \, W^{\mu \, \nu} \, W_{\mu \, \nu}.
\eea
Note that some notation is reused here from the SMEFT operator basis.
Decays through the {latter} operator lead to 
enhanced couplings to $W^+ W^-$, $ZZ$ and $Z \, \gamma$ final 
states, while the {first two yield smaller
levels of}
such decays. Decays to $Z \gamma$ 
are disfavoured by correlated searches at the mass scale $\sim 750 \, {\rm GeV}$. Ref \cite{Aad:2014fha} reports a $95 \%$ C.L. bound on 
$\sigma(pp \rightarrow S \rightarrow Z(\ell^+ \, \ell^-) \, \gamma)$ of $\lesssim 0.3
\, {\rm fb}$, whereas the expected  
 deviation {associated with} the
$\gamma \gamma$ excess, assuming the decay is generated by 
{the coupling to $\rm SU(2)_L$ (and the $Z$ and decaying to $\ell^+ \, \ell^-$)}, is 
$\sim 2 \, \rm{fb}$.\footnote{Here we have used $\sigma (pp \rightarrow s)[8 \, {\rm TeV}]/\sigma (pp \rightarrow s)[13 \, {\rm TeV}] \cong 0.21$,
assuming $gg$ production is dominant. Note that the excess is in 
13 {\rm TeV} data while the bound in Ref.\ \cite{Aad:2014fha} 
is for 8 {\rm TeV} data.} It is possible to {tune away} this tension
by having both the operators $O_B$ and $O_W$ present with 
correlated Wilson {coefficients} 
\cite{Harigaya:2015ezk} (implicitly defined in Eqn.\ref{basicL}); however 
we will not {further} consider generating $c_W$ {since} $Z \gamma$ bounds 
are not the essential point {of this
study}.
The decay widths are related to the remaining dimension-five 
operators, introduced with the given normalization, as
\bea
\label{Gammas}
\Gamma(S \rightarrow \gamma \, \gamma) = \frac{4 \, \pi \, \alpha_{ew}^2 \, m_s^3}{\Lambda_{\gamma}^2} \, c_B^2,  \quad \Gamma(S \rightarrow gg) = \frac{32 \, \pi \, \alpha_{s}^2 \, m_s^3}{\Lambda_{g}^2} \, c_G^2.
\eea

Using (\ref{Gammas}), we can rewrite the cross section for the 
diphoton excess (\ref{delta_sigma}) as 
\bea
\label{dsigma2}
\frac{\Delta \sigma(pp \rightarrow  \gamma \, \gamma)}{8 [fb]}\left(\frac{\Gamma_s}{45 \, {\rm GeV}} \right) \cong 6546 \, \left(\frac{m_s^2 \, c_B^2}{\Lambda_\gamma^2}\right)  \, \left[\left(\frac{m_s^2 \, c_G^2}{\Lambda_g^2} \right) +  2.4 \times 10^{-7} \, \left(\frac{m_s^2 \, c_B^2}{\Lambda_\gamma^2} \right)  \right].
\eea
The gauge couplings $\alpha_s,\alpha_{ew}$ are evaluated 
at the scale $m_s \cong 750 \, {\rm GeV}$. We note that,
in the presence of the operators generated by integrating out 
the scalar $S$ at its mass, 
the running of $\alpha_s,\alpha_{ew}$ is modified \cite{Jenkins:2013zja}. The corresponding Wilson coefficients in the SMEFT can receive contributions from other unknown UV physics.  Such nonresonant contributions are neglected. We also note that 
the running effect on the production and decay of the scalar particle is higher order in the power counting, and neglected.
We run $\alpha_s,\alpha_{ew}$  up from the scale $m_Z$ using SM relations,
so that $ \alpha_s(750 \, {\rm GeV}) \cong 0.09,$ and $\alpha_{ew}(750 \,{\rm GeV}) \cong 1/126.5$.

In a valid EFT expansion, one expects that
$m_s < \Lambda_{\gamma,g}$. If we normalized the Wilson coefficients proportional to a loop factor $\sim (16\pi)^{-2}$
in the case of some weakly coupled renormalizable UV models,  large Wilson coefficients are required.
Extreme solutions where $c_B/\Lambda_\gamma \ll c_G/\Lambda_g$ or $c_B/\Lambda_\gamma \gg c_G/\Lambda_g$ are possible. One naturally expects $\Lambda_g \sim \Lambda_\gamma$
and $c_B$ and $c_G$ to differ only by group theory factors, in scenarios where a common mediator generates the two decays.

\subsection{Integrating out $S$}
Minimal scalar field models have the potential Lagrangian terms
\bea
\mathcal{L}_V &=&  -\lambda_{SM} \left(H^\dagger H -\frac12 v^2\right)^2 - \frac{m_s^2}{2} \,S^2  +  \frac{\kappa}{4!} \, S^4
+  \, \lambda \, \Lambda_c \, S \, H^\dagger H +  \lambda_2 \, \Lambda_c \, S^3 +  \lambda_3 \, S^2\, (H^\dagger H) + \cdots \nn
\eea
No unbroken discrete symmetry exists that forbids the $\lambda, \lambda_2$ terms, since $S$ decays.
Here $\Lambda_c$ is the cutoff scale of the toy model effective Lagrangian, and we assume that some unknown states with a mass scale $\sim \Lambda_c$
generate the coupling of $S$ to photons and gluons. 
Due to the presence of effective dimension five terms in the Lagrangian, higher order terms are also generated in the potential suppressed by $1/\Lambda_c$.
Since $m_s \gg v$, it is interesting to consider the case that $S$ decays through manifestly $\rm SU(3) \times SU(2)_L \times U(1)_Y$ invariant operators.
Integrating out $S$
one obtains the effective lagrangian suitable for describing Higgs-gauge boson couplings,
\bea
\mathcal{L} = \mathcal{L}_{SM} + \frac{C_{HG}(m_s) \, g_3^2}{\Lambda^2} H^\dagger \, H \, G^{\mu \, \nu} \, G_{\mu \, \nu} + \frac{C_{HB}(m_s) \, g_1^2}{\Lambda^2} H^\dagger \, H \, B^{\mu \, \nu} \, B_{\mu \, \nu},
\eea
where 
\bea\label{eq:match}
\frac{C_{HG}(m_s)}{\Lambda^2} =  \frac{c_G \, \lambda}{m_s^2} \, \frac{\Lambda_c}{\Lambda_g}, \quad \frac{C_{HB}(m_s)}{\Lambda^2} =  \frac{c_B \, \lambda}{m_s^2}\,  \frac{\Lambda_c}{\Lambda_\gamma}.
\eea
At lower scales $\mu$, the Wilson coefficients are then, in a leading log approximation, (only retaining the Yukawa couplings $Y_t,Y_b$) \cite{Jenkins:2013zja,Jenkins:2013wua,Alonso:2013hga,Alonso:2014zka}
\bea
C_{HG}(\mu) &=& \left(1- \log\left[\frac{{m_s}}{\mu}\right] \, \frac{12 \, \lambda + 2 \, N_c \, ((\sqrt{2} M_t)^2+ (\sqrt{2} M_b)^2)/v^2 - 6 \, g_1^2 y_h^2 - 9 \, g_2^2/2}{16 \pi^2}\right) C_{HG}(m_s), \nonumber \\
C_{HB}(\mu) &=& \left(1- \log\left[\frac{{m_s}}{\mu}\right] \, \frac{12 \, \lambda + 2 \, N_c \, ((\sqrt{2} M_t)^2+(\sqrt{2} M_b)^2)/v^2 + 2 y_h^2 g_1^2 - 9 \, g_2^2/2}{16 \pi^2}\right) C_{HB}(m_s),  \nonumber \\
C_{uH}(\mu) &=& - \frac{\sqrt{2} M_t}{v} \, \frac{2  \, g_3^4}{\pi^2} \, \log\left[\frac{{m_s}}{\mu}\right] \, C_{HG}(m_s) - \frac{\sqrt{2}M_t}{v}\frac{3(y_h^2 + 2 y_q y_u)}{4 \pi^2} g_1^4\log\left[\frac{{m_s}}{\mu}\right] C_{HB}(m_s), \nonumber \\
C_{dH}(\mu) &=& - \frac{\sqrt{2} M_b}{v} \, \frac{2  \, g_3^4}{\pi^2} \, \log\left[\frac{{m_s}}{\mu}\right] \, C_{HG}(m_s) - \frac{\sqrt{2}M_b}{v}\frac{3(y_h^2 + 2 y_q y_d)}{4 \pi^2}g_1^4\log\left[\frac{{m_s}}{\mu}\right] C_{HB}(m_s), \nonumber \\
C_{uG}(\mu) &=& \frac{\sqrt{2} M_t}{4 \, \pi^2 \, v}  g_3^3  \, \log\left[\frac{{m_s}}{\mu}\right] C_{HG}(m_s), \nn \\
C_{uB}(\mu) &=&  \frac{\sqrt{2} M_t}{16 \, \pi^2 \,v}  (2 g_1^3 \, ( y_q +  y_u)) \,\log\left[\frac{{m_s}}{\mu}\right] \,  C_{HB}(m_s), \nn \\
C_{dG}(\mu) &=& \frac{\sqrt{2} M_b}{4 \, \pi^2 \, v}  g_3^3  \,  \, \log\left[\frac{{m_s}}{\mu}\right] \,C_{HG}(m_s), \nn \\
C_{dB}(\mu) &=&  \frac{\sqrt{2} M_b}{16 \, \pi^2 \, v}  (2 g_1^3 \, ( y_q +  y_d)) \, \log\left[\frac{{m_s}}{\mu}\right] \,C_{HB}(m_s), \nn \\
C_{HWB}(\mu)&=& - \frac{4 g_1^3 g_2 y_h  }{16 \pi^2}  \log\left[\frac{{m_s}}{\mu}\right]C_{HB}(m_s),
\eea
with $y_h=1/2, y_q=1/6, y_u=2/3, y_d=-1/3$ being the hypercharges of the indicated particles. Here the $O_{uH},O_{dH},O_{eH},O_{uG}$, $O_{dG},O_{uB}$, $O_{dB},O_{HWB},O_H$ operators are defined as in Ref.\cite{Grzadkowski:2010es}. The Higgs potential is also changed in a nontrivial fashion
\bea
\delta \lambda_{SM}(\mu) &=& - \frac{3 m_h^2}{4\pi^2} \, g_1^4 \, y_h^2 \,  \log\left[\frac{{m_s}}{\mu}\right] \,C_{HB}(m_s),  \\
C_H(\mu) &=& - \frac{3}{\pi^2} \, \lambda \left(g_1^4 \, y_h^2 \right) \,\log\left[\frac{{m_s}}{\mu}\right] \, C_{HB}(m_s) + \left(\frac{3}{\pi^2} g_1^6 \, y_h^4 + \frac{3}{4\, \pi^2} g_1^4 \, g_2^2 \, y_h^2\right) \,\log\left[\frac{{m_s}}{\mu}\right] \, C_{HB}(m_s). \nn
\eea
Here $\delta \lambda_{SM}$ is the modification of the running of the Higgs self-coupling relative to how it runs in the SM, below the scale $m_s$.
It is interesting to note that these one-loop effects would have a nontrivial implication for the running and shape of the potential, if the diphoton excess was substantiated.
We will resist drawing conclusions about the fate of the universe due to this observation. The modified potential will redefine the effective Higgs vacuum expectation value,
but in an unobservable fashion in current experiments.
For the particular case of Higgs physics, to get a sense of the impact on lower energy phenomenology, we note
\bea
C_{HG}(m_h) &\cong& 0.94 \, C_{HG}(m_s), \\
C_{HB}(m_h) &\cong& 0.94 \, C_{HB}(m_s),  \\
C_{uH}(m_h) &\cong& -0.46 \, C_{HG}(m_s)  - 0.0011 \, C_{HB}(m_s), \\
C_{dH}(m_h) &\cong& -0.011 \, C_{HG}(m_s)  - 7.85 \times 10^{-6} \, C_{HB}(m_s), \\
C_{uG}(m_h) &\cong& 0.054 \, C_{HG}(m_s), \\
C_{dG}(m_h) &\cong& 0.0013 \, C_{HG}(m_s), \\
C_{uB}(m_h) &\cong& 8.8 \times 10^{-4} \, C_{HB}(m_s), \\
C_{dB}(m_h) &\cong& -4.4\times 10^{-6} \, C_{HB}(m_s), \\
C_{HWB}(m_h) &\cong& -6.9 \times 10^{-4}\,C_{HB}(m_s).
\eea
Flavour indices have been suppressed, due to the scenario considered. There is a (up-down quark) flavour non-universal effect.
Note the large effect on the top Yukawa coupling at the low scale.
The (assumed) SM Higgs field at 125 {\rm GeV} coupling to the top gets modified as
\bea
\kappa_t -1 =  \frac{v^2}{\sqrt{2} \, \Lambda^2} \left(0.46 \, C_{HG}(m_s)  + 0.0011 \, C_{HB}(m_s)\right).
\eea
Similarly one finds a modification for the coupling of the $h \, Z \, Z$  interaction of the form
\bea
(0.94 s_\theta^2 ) \, \sqrt{2} \, v \, h \, Z^{\mu \, \nu} \, Z_{\mu \, \nu} \, C_{HB}(m_s)
\eea
with $s_\theta^2$ referring to the Weinberg angle. The correction to the angle due to $C_{HWB}$ for this term is higher order.
This correction leads to an effective modification of $\kappa_Z$. Taking into account the typical offshellness in the decays $h \rightarrow Z \, Z^\star$ into
fermion final states \cite{Contino:2013kra}, one finds
\bea
\kappa_Z -1 \cong 0.2 \, \frac{m_w^2}{\Lambda^2} C_{HB}(m_s).
\eea
The effects on $\kappa_Z$, $\kappa_t$ are in general subdominant in the minimal scenario considered and can be neglected.
In general, a scalar singlet of the form considered above is well isolated from inducing large low-energy effects.

As states of mass scale $\Lambda_c$ generate  the $c_B$ operator at the scale $\Lambda_\gamma$,
it is necessary that they are charged under $U(1)_Y$. One expects a large number of operators to be generated at the scale
$\Lambda_c$, with contributions to operators that include SM states that are charged under $U(1)_Y$, at least at the two-loop level.
In this case, the detailed impact on low-energy phenomenology can differ from the minimal case sketched here. 
When the Wilson coefficient of $S \, H^\dagger \, H$ is suppressed, two-loop effects can be comparable, or dominant, over
the effects that we study in detail. One also expects one-loop contributions to the operators $O_{HG},O_{HB}$ on general grounds. Our analysis assumes that such direct matching contributions are small enough to be neglected.
The couplings of the states that generate $C_{G,B}$ to the SM Higgs are unknown and can be small.

\subsection{Generating the $S \, H^\dagger \, H$ operator}\label{sec:loopmixing}

On general grounds one expects the coupling of the scalar state to be sizable with $H^\dagger \, H$, and for the scalar to have a sizable self-coupling term
$S^3$. These operators are relevant.  
Of course, pure naturalness expectations for scalar sectors are under pressure due to the measured Higgs mass.
Assuming $\lambda \, \Lambda_c \sim [\rm TeV]$ the induced Higgs mass value is not strongly perturbed as $\Delta m_h^2 \sim [\rm TeV]^2/8 \pi^2$,
so considering separations of scales where $\lambda \, \Lambda_c > m_s$ by an order one factor does not introduce significant extra tuning to the Higgs sector.

As we will characterize in more detail below, the scenario where $\lambda \, \lambda_c$ is a value proximate to the cutoff scale, is problematic.\footnote{This observation was also pointed
out while this draft was being finalized in Ref \cite{Falkowski:2015swt}.}
At scales below $m_s$,  the presence of the $S \, H^\dagger \, H$ operator leads to the higher dimensional operators $O_{HB}, O_{HG}$ in the SMEFT. 
If it is assumed that $\lambda \, \lambda_c$ is somehow suppressed, or fixed to zero, quantum corrections regenerate this mixing due to the
interactions assumed above, to explain the excess in Eqn.\ref{basicL}. The mixing between $S$ and $h$ due to $c_B$ is
generated by the one-loop diagram shown in Fig.1 a. This gives a Naive Dimensional Analysis (NDA) \cite{Manohar:1983md} estimate of the coefficient of $S \, H^\dagger \, H$
of the form 
\bea\label{SHHtree:mixing}
 \lambda \, \Lambda_c \gtrsim \frac{g_1^4 \, m_s^2}{32 \, \pi^2} \, \frac{c_{B}(m_s)}{\Lambda_\gamma},
\eea
when the bare and loop induced terms of  the Wilson coefficient of the operator $S H^\dagger \, H$ are not canceled against one another, at the scale $\mu = m_s$. 
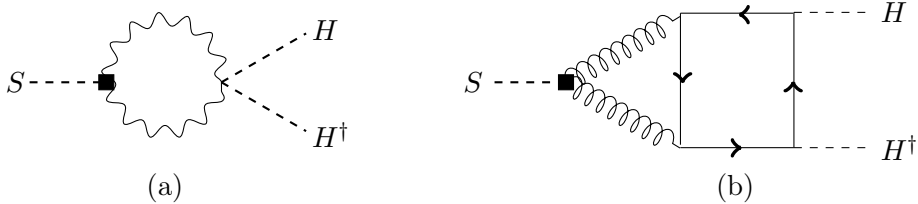
\begin{figure}[!h]
\hspace{2cm}
\begin{tikzpicture}
\draw  [decorate,decoration=snake] (0+3,0) circle (0.75);

\filldraw (-0.68+3,-0.1) rectangle (-0.88+3,0.1);

\draw[dashed] [thick] (0:0.75+3) -- +(30:1.30) ;
\draw[dashed] [thick] (0:0.75+3) -- +(330:1.30) ;

\draw  [dashed] [thick] (-1.8+3,0) -- (-0.75+3,0);

\node [left][ultra thick] at (-1.7+3,0) {$S$};
\node [right][ultra thick] at (1.80+3,0.7) {$H$};
\node [right][ultra thick] at (1.80+3,-0.7) {$H^{\dagger}$};
\draw (3,-1.4) node [align=center] {(a)};
\end{tikzpicture}
\hspace{1cm}
\begin{tikzpicture}
\draw  [dashed] [thick] (-1.7,0) -- (-0.75,0);
\filldraw (-0.65,-0.1) rectangle (-0.85,0.1);

\draw[decorate,decoration={coil,amplitude=4pt, segment length=5pt}] (180:0.75) -- +(30:1.75) ;
 
 \draw[decorate,decoration={coil,amplitude=4pt, segment length=5pt}] (180:0.75) -- +(330:1.75) ;

\draw (50:1.2) -- +(270:1.75) ;
\draw  [->][ultra thick]  (0:0.8)  -- + (270:0.01) ;

\draw (50:1.2) -- +(0:1.5) ;
\draw  [->][ultra thick]  (0:2.27)  -- + (90:0.01) ;
\draw  [->][ultra thick]  (-29:1.8)  -- + (0:0.01) ;
\draw  [->][ultra thick]  (30:1.8)  -- + (180:0.01) ;
\draw (0:0.8)  -- + (270:0.01);
\draw (2.28,-0.87) -- (2.28,0.93);
\draw (0.77,-0.87) -- (2.28,-0.87);

\draw [dashed] (2.28,0.93) -- (3.28,0.93);
\draw [dashed] (2.28,-0.87) -- (3.28,-0.87);

\node [left][ultra thick] at (-1.7,0) {$S$};
\node [right][ultra thick] at (3.28,0.93) {$H$};
\node [right][ultra thick] at (3.28,-0.87) {$H^{\dagger}$};
\draw (1.5,-1.4) node [align=center] {(b)};

\end{tikzpicture}
\caption{\label{fig:1} 
Diagrams generating the mixing of $S$ and $H$ at one loop due to the couplings required for $gg \rightarrow S$ and $S \rightarrow \gamma \, \gamma$,
illustrated with the insertion of a box.}
\end{figure}
The generation of $S \, H^\dagger \, H$ due to $c_G$ is a two-loop effect. A typical diagram is given in Fig.1 b. The divergence in the diagram
leading to the mixing is approximately 
\bea
\sim  {\rm Tr} \left[T_A T_A\right] y_t^2 g_3^2 \frac{m_s^2 \, c_G}{(16 \pi^2)^2 \epsilon \Lambda_g},
\eea
in dimensional regularization in $d = 4 - 2 \epsilon$ dimensions. The colour factors enhance the magnitude of the diagram, as expected\footnote{Note that this is only a single pole divergence, despite being a two-loop graph. This is because the subgraph coupling $H^\dagger H$ to two gluons is finite in the SM}.
The $\epsilon$ poles cancel in the matching onto the lower energy theory. However, we  utilize the corresponding
finite terms generated from the logarithmic dependence linked to the divergence in this diagram as an NDA-inspired estimate of the size of the Wilson coefficient of the $S H^\dagger \, H$ operator.
We give the exact result for the pole of the sum of the diagrams of this form in the Appendix.
\subsection{Direct matching contributions to the $O_{HB}$, $O_{HG}$ operators}\label{sec:matching}

There are also other direct one-loop
contributions to the Wilson coefficients of the SMEFT operators shown in Fig.2 a,b. Fig.2 a depends on an unknown scalar coupling in the potential -- $\lambda_3$,
but does not require the operator $S H^\dagger \, H$ to generate an effective low scale $C_{HB}$,$C_{HG}$.
Consider calculating Fig.2 a in dimensional regularization.
The matching coefficient onto the SMEFT then receives a contribution from the finite parts of the on shell diagrams in the full and effective theories (in this case the SMEFT), while dropping the $1/\epsilon$ poles.\footnote{
In the full and effective theory, the UV poles cancel, and the IR poles are the same between the two theories by definition. See Ref.\cite{Manohar:2003vb} for more discussion.} We have calculated the diagrams in Fig.2 (see the Appendix); a simple 1-loop estimate of the NDA minimum for the SMEFT operator's Wilson coefficients is adequate for our bounds. We require that
\bea
|\frac{C_{HB}}{\Lambda^2} |&\gtrsim&\frac{\left(\lambda_3 + y^2_h \, g_1^2 \right) g_1^2 \, c_B^2}{4 \pi^2 \, \Lambda_\gamma^2},\\
|\frac{C_{HG}}{\Lambda^2}|&\gtrsim&\frac{\lambda_3 \, g_3^2 \, c_G^2}{4 \, \pi^2 \, \Lambda_g^2}.
\eea

The contributions to $C_{HB}$ depend on different combinations of unknown parameters in the UV theory. They are expected to not be simultaneously tuned to be small in "natural" scenarios. The $C_{HG}$ contribution only proceeds through the scalar quartic interaction.
Interestingly, Fig. 2a,b do not vanish in the case of pseudoscalar effective operators; we will return to this point in Section \ref{pseudo}.
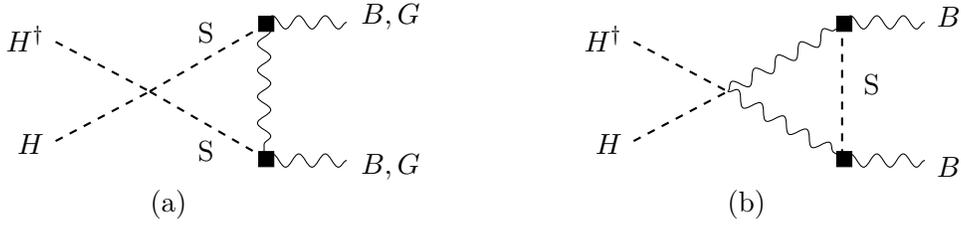
\begin{figure}[!h]
\begin{tikzpicture}
\hspace{1.2cm}
\draw  [dashed] [thick] (-2,0.66) -- (-0.75,0);
\draw [dashed] [thick] (-2,-0.66) -- (-0.75,0);

\draw[thick] [dashed] (180:0.75) -- +(30:1.75) ;
\draw[thick] [dashed] (180:0.75) -- +(330:1.75) ;
\draw[decorate,decoration=snake] (50:1.2) -- +(270:1.75) ;

\draw[decorate,decoration=snake] (50:1.2) -- +(0:1.1) ;
\draw[decorate,decoration=snake] (310:1.2) -- +(0:1.1) ;

\node [left][ultra thick] at (-2,0.7) {$H^{\dagger}$};
\node [left][ultra thick] at (-2,-0.7) {$H$};
\node [left] [ultra thick] at (3,1) {$B,G$};
\node [left] [ultra thick] at (3,-1) {$B,G$};
\node [above] [ultra thick] at (0,0.5) {S};
\node [below] [ultra thick] at (0,-0.5) {S};

\filldraw (0.7,0.8) rectangle (0.9,1);
\filldraw (0.7,-0.8) rectangle (0.9,-1);
\draw (-0.5,-1.5) node [align=center] {(a)};
\end{tikzpicture}
\begin{tikzpicture}
\hspace{3cm}
\draw  [dashed] [thick] (-2,0.66) -- (-0.75,0);
\draw [dashed] [thick] (-2,-0.66) -- (-0.75,0);

\draw[decorate,decoration=snake] (180:0.75) -- +(30:1.75) ;
\draw[decorate,decoration=snake] (180:0.75) -- +(330:1.75) ;
\draw[thick][dashed] (50:1.2) -- +(270:1.75) ;

\draw[decorate,decoration=snake] (50:1.2) -- +(0:1.1) ;
\draw[decorate,decoration=snake] (310:1.2) -- +(0:1.1) ;

\node [left][ultra thick] at (-2,0.7) {$H^{\dagger}$};
\node [left][ultra thick] at (-2,-0.7) {$H$};
\node [left] [ultra thick] at (2.5,1) {$B$};
\node [left] [ultra thick] at (2.5,-1) {$B$};
\node [right] [ultra thick] at (0.9,0.1) {S};

\filldraw (0.7,0.8) rectangle (0.9,1);
\filldraw (0.7,-0.8) rectangle (0.9,-1);
\draw (-0.5,-1.5) node [align=center] {(b)};
\end{tikzpicture}
\caption{\label{fig:1} 
Direct matching diagrams to $O_{HB}$,$O_{HG}$ at one loop due to the couplings required for $gg \rightarrow S$ and $S \rightarrow \gamma \, \gamma$,
illustrated with the insertion of a box. Note that these diagrams are also generated by two insertions of the operators with $\tilde{B}_{\mu \, \nu} B^{\mu \, \nu}$ and
$\tilde{G}_{\mu \, \nu} G^{\mu \, \nu}$. These operators are present in the pseudoscalar case. We discuss this case in Section \ref{pseudo}. Note that a box diagram contribution of this form is not shown as it
vanishes due to Lorentz index interchange symmetry.}
\end{figure}

\subsection{Constraints from Electroweak Precision Data}
In \cite{Berthier:2015gja}, a global fit in the SMEFT has been performed incorporating data from PEP, PETRA, TRISTAN, SpS, Tevatron, SLAC, LEPI and LEPII. Bounds on a number of Wilson coefficients have been obtained and theoretical errors in the SM as well as in the SMEFT have been studied and included, which leads to a relaxation of these bounds. Among these Wilson coefficients one ($C_{HWB}$ - also known as the $S$ parameter) is of particular interest as it is generated by $C_{HB}$ by its running from the higher energy scale $\sim m_s$ \cite{Hagiwara:1992eh,Alam:1997nk,Grojean:2013kd}. 
All other Wilson coefficients not generated by the running of $C_{HB}$ and $C_{HG}$ are set to zero in the fit, allowing us using the same data as in \cite{Berthier:2015gja}, to put constraints on $C_{HWB}$ at a low-energy scale $m_Z$. This can be translated into bounds on $C_{HB}$ at $m_s$ using the RGE for $C_{HWB}$ for which we take $C_{HWB}(m_s) \simeq 0$.
The other Wilson coefficients are not exactly zero in any realistic model, but are assumed subdominant. We introduce a theoretical error for the SMEFT to take this into account consistently. 

We give the best fit value $\tilde{C}_{HWB}^{min}(m_Z) \pm \sigma$ with $\tilde{C}_{HWB} = 100 C_{HWB}$ as well as resulting bounds on $C_{HB}(m_s)$ for a SMEFT error $=\{ 0\%, 0.5 \%, 1\% \}$ in Tab.~\ref{EWPD}. Using the RGE of $C_{HWB}$, $C_{HB}(m_s) = C_{HB}^{min}(m_s) \pm \sigma^{'}$ with $C_{HB}^{min}(m_s) = - C_{HWB}^{min}(m_Z)/ K$, $\sigma^{'} = \sigma/ K$ and $K$ given by
\bea
K= \frac{4 g_1^3 g_2 y_h}{16 \pi^2}\log\left[\frac{m_s}{m_Z}\right].
\eea

\begin{center}
\begin{table}[h]
\centering
\tabcolsep 8pt
\begin{tabular}{|c|c|c|c|}
\hline
SMEFT error & $0\%$& $0.5 \%$&$1 \%$ \\
\hline 
$\tilde{C}_{HWB} (m_Z)v^2/\Lambda^2$&$-0.0097 \pm 0.018$ & $0.024 \pm 0.028$ & $0.018 \pm 0.030$ \\
\hline
$C_{HB} (m_s)v^2/\Lambda^2$&$0.12 \pm 0.23$&$-0.29 \pm 0.35 $&$-0.23 \pm 0.37 $ \\
\hline\end{tabular}
\caption{Bounds on $\tilde{C}_{HWB}(m_Z)v^2/\Lambda^2$ and the resulting bounds on $C_{HB}(m_s)v^2/\Lambda^2$ for a SMEFT error = $\{ 0\%, 0.5 \%, 1\% \}.$ \label{EWPD}}
\end{table}
\end{center}
Neglecting the running of $C_{HB}$ between the energy scales $m_s$ and $m_h$, we can use the EWPD to extract bounds on $\kappa_{\gamma}$. We quote the bounds obtained on $\kappa_{\gamma}$ in Tab.~\ref{kappa gamma} which are very weak.
\begin{center}
\begin{table}[h]
\centering
\tabcolsep 8pt
\begin{tabular}{|c|c|c|c|}
\hline
SMEFT error & $0\%$& $0.5 \%$&$1 \%$ \\
\hline
$\kappa_{\gamma}-1 $&$+12 \pm 23 $&$-29 \pm 34$&$-22 \pm 37$ \\ 
\hline\end{tabular}
\caption{Bounds on $\kappa_{\gamma}$ from EWPD for a SMEFT error = \{ 0 \%, 0.5 \%, 1 \%\} \label{kappa gamma} in this minimal scenario.}
\end{table}
\end{center}

\subsection{Constraints from {run I} Higgs data}
The operators $O_{HB}$ and $O_{HG}$ map to the $\kappa_g$ and $\kappa_\gamma$ parameters as
\bea
\kappa_g = 1 - \frac{16 \, \pi^2 v^2 \, C_{HG}(m_h)}{\Lambda^2 \, I_g}, \quad   \quad  \kappa_\gamma = 1 - \frac{16 \, \pi^2 v^2 \, C_{HB}(m_h)}{\Lambda^2 \, I_\gamma}.
\eea
Here we are using notation consistent with Ref.\cite{Manohar:2006gz}. As the $\kappa$ couplings are defined with respect to rescaling the best SM predictions,
we retain the NLO QCD correction in the heavy top limit in the expressions $I_g, I_\gamma$ quoted in Ref.\cite{Manohar:2006gz}. We neglect a correction due to known NLO EW terms
that are included in the scaled out SM value experimentally. This introduces an error on the order of $\alpha_{ew}/4 \, \pi \, v^2/\Lambda^2$.
We use $I_g \simeq 0.38, I_\gamma \simeq  -1.6$, retaining only the top quark contribution to the loop functions for the fermions.

In the minimal predictive scenario considered so far, the modified top coupling $\kappa_t$ is related to $\kappa_\gamma$ and $\kappa_g$ as
\bea
\kappa_t -1 =  - \frac{v^2}{\sqrt{2}} \left(0.46 \, \frac{(\kappa_g-1) I_g}{16 \, \pi^2}  + 0.0011 \, \frac{(\kappa_\gamma-1) \, I_\gamma}{16 \, \pi^2}\right).
\eea 
so that it is justified to neglect $\kappa_t$ as sub-leading and consider the constraints from global Higgs data analyses in just the $(\kappa_g, \kappa_\gamma)$ space. 
So far our discussion has been general.
 \subsubsection{Mixing Domination}

The tension with the measurements of  $\kappa_g,\kappa_\gamma$ reported for the 125 {\rm GeV} scalar, when mixing is assumed to dominate the contribution
to the low-energy phenomenology through the operator $S \, H^\dagger \, H$, can be characterized by the parameter $\mathcal{O}$ defined as
\bea\label{param}
\mathcal{O} = \frac{\Delta \sigma(pp \rightarrow S \rightarrow \gamma \, \gamma)}{8 [fb]}\left(\frac{\Gamma_s}{45 \, {\rm GeV}} \right)  \left(\frac{\lambda \,\Lambda_c}{ N \times 750\, {\rm GeV}} \right)^4,
\eea
which is expected to be order one.
Here $N$ is a factor for the separation of the cutoff scale and $m_s$. By definition $\Lambda_c \gtrsim m_s$, and we take $N=3$ below.
The measured excess leads to the constraint on $\kappa_g, \kappa_\gamma$
\bea\label{main:constraint}
\mathcal{O}  \simeq 0.0005  \, \left(\kappa_\gamma -1 \right)^2 \left[\left(\kappa_g -1 \right)^2 + 4.2 \times 10^{-6} \, \left(\kappa_\gamma -1 \right)^2 \right]. 
\eea 
The deviations $|\kappa_g-1|,|\kappa_\gamma-1|$ are constrained to be $ \lesssim 0.25$ at $95 \%$ C.L \cite{ATLAS-CONF-2015-044}. 
We illustrate this relation in Fig.3.
\begin{figure}[t]
  \centering
  \includegraphics[width=3.0in,height=3in]{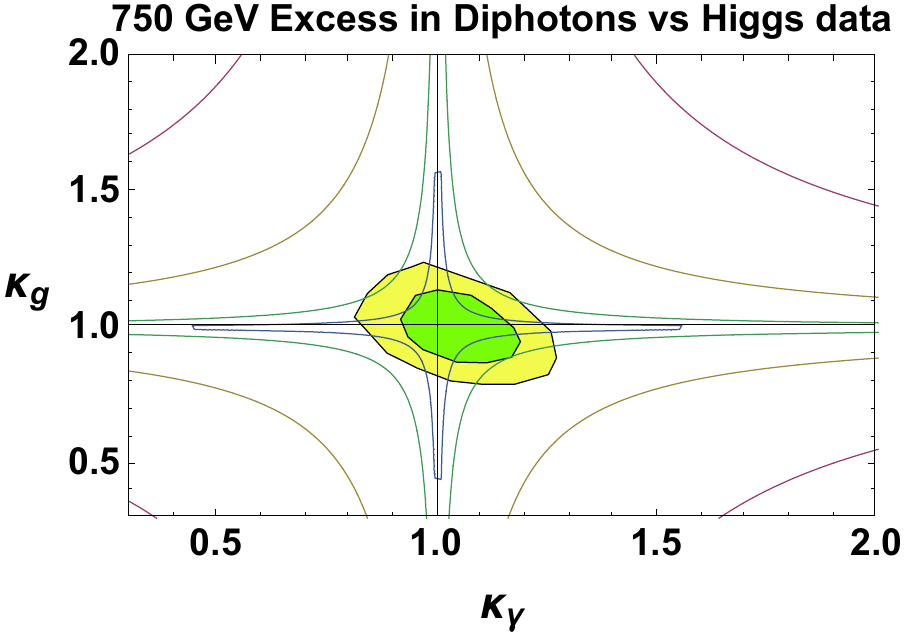}
    \includegraphics[width=3.0in,height=3in]{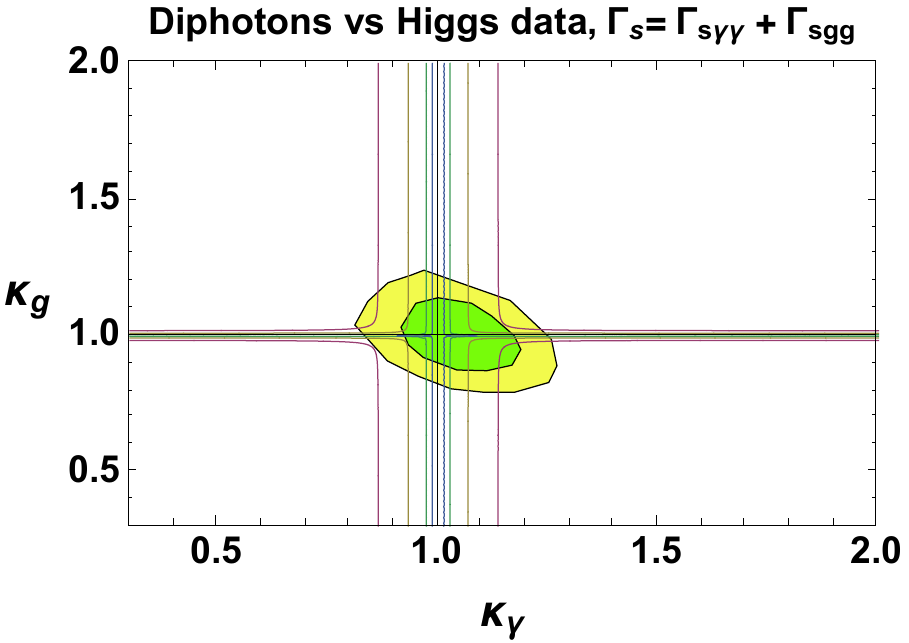}
  \caption{An illustration of the tension between Higgs data and the diphoton excess in minimal scalar models. The curves in the left hand plot are values of $\lambda \Lambda_c/N \times 750 \, {\rm GeV}  = \{0.1,0.05,0.02,0.01\}$ 
  coming in from the outermost curve in Eqn.\ref{main:constraint}. The right hand plot shows $\lambda \Lambda_c/N \times 750 \, {\rm GeV}   = \{0.1,0.05,0.02,0.01\}$
  in the case that the width is reproduced by just the gluonic and photon production and decay, given by Eqn.\ref{widthfix}. In both figures, the curves are overlaid on the
  $68\%$ and $95\%$ CL curves from the {run I} Atlas-CMS Higgs combination, fitting only to $\kappa_g, \kappa_\gamma$ \cite{ATLAS-CONF-2015-044}.}
  \label{Fig;nsvsr}
\end{figure}
This conflict can be
relaxed in a linear fashion if the excess decreases from its reference value of $8 [fb]$ or the width decreases from its reference value of $45 \, {\rm GeV}$.
However, the inconsistency for order one mixing angles is at the level of four orders of magnitude. The coupling of $S$ to $H^\dagger \, H$ that scales as a fourth power must be suppressed
from "natural" values to restore consistency with {run I} data. By the same token, the suppression does not have to be dramatic. An order of magnitude to the
fourth power in suppression makes the scenario consistent, considering the experimental uncertainties on the small excess at $750 \, {\rm GeV}$. Two orders of magnitude suppression in the coupling
of coupling of $S$ to $H^\dagger \, H$ restores good agreement with low-energy Higgs data, and such a suppression is not strongly challenged by naturalness concerns. 
\subsubsection{Reproducing the Width}
If we further fix the condition that $\Gamma_s = \Gamma(s\rightarrow \gamma \,\gamma)+ \Gamma(s\rightarrow gg)$, we derive the constraint equation to reproduce the excess
\bea\label{widthfix}
\frac{\Delta\sigma\left(pp\to S\to\gamma\gamma\right)}{8[fb]}\left(\frac{\lambda\Lambda_c}{Nm_s}\right)^2= 0.06\,N^2 \, \left(\kappa_\gamma-1\right)^2\frac{\left(\kappa_g-1\right)^2+ 4.2 \times10^{-6}\left(\kappa_\gamma-1\right)^2}{\left(\kappa_g-1\right)^2+0.017\left(\kappa_\gamma-1\right)^2}
\eea
The effects of this condition are shown in Fig.3. Note that reducing the width in this case quickly allows consistency with lower energy data, by making the coupling required to reproduce the excess smaller.

 \subsubsection{Matching Domination}\label{sec:matchbounds}
As we have stressed, the Wilson coefficients $C_{HB}$ and $C_{HG}$ also receive contributions independent of the mixing angle. As these matching coefficients are generated by loops involving two insertions of the new scalar's coupling to SM field strengths, they lead to a relation between the measured excess and Higgs data which scales as just a square rather than a fourth power.

In the limit where these matching contributions are the only contribution to the shifts in the Higgs couplings and $\lambda_3\gg\frac{y_h^2 \,g_1^2}{2}$, we can express the signal rate as

\bea
\frac{\Delta \sigma(pp \rightarrow S \rightarrow \gamma \, \gamma)}{8 [fb]}\left(\frac{\Gamma_s}{45 \, {\rm GeV}} \right)=\frac{1.7\times10^5}{\lambda_3^2}\left(\kappa_\gamma-1\right)\left(\left(\kappa_g-1\right)+1.5\times10^{-4}\left(\kappa_\gamma-1\right)\right). \nonumber \\
\eea

These matching contributions to the Higgs observables are not significantly constrained by the {run I} Higgs data.

\section{Consistency of pseudoscalar models with lower energy data}\label{pseudo}

A $J^P = 0^-$ pseudoscalar boson  interpretation of the $S$ particle related to the diphoton excess would not lead to direct mix with the $J^P = 0^+$ Higgs boson. 
This further protects this model from related low-energy phenomenology constraints. 
At the one-loop level such interactions still generate $H^\dagger H B^{\mu \, \nu} \, B_{\mu \, \nu}$ through the diagrams shown in Fig.2. We have calculated these contributions and found them to be identical to the scalar case; therefore, the discussion in Section \ref{sec:matchbounds} applies in full to pseudoscalar models. This leaves models which employ a fundamental pseudoscalar to explain the diphoton excess largely unconstrained.

However, the constraints on mixing
discussed above still apply to a heavy sector with such a state, which generally arise when the pseudoscalar being considered is a bound state of new strong dynamics. To elaborate on this point concretely, we 
utilize the models discussed in Ref. \cite{Harigaya:2015ezk}. Consider a minimal "hidden pion" model of a pseudoscalar given in Ref. \cite{Harigaya:2015ezk},
which also introduces heavy vector-like hidden quarks at the scale $\Lambda_c$, and a new $SU(N)$ gauge group.
This leads to the effective interactions of the "hidden pion" $\phi$
\bea
\mathcal{L}_\phi = c_{\tilde{G}}^\phi \, \frac{\phi}{f} \, G^{\mu \, \nu} \, \tilde{G}_{\mu \, \nu} + c_{\tilde{B}}^\phi  \, \frac{\phi}{f} \, B^{\mu \, \nu} \, \tilde{B}_{\mu \, \nu}
\eea
with $\tilde{G}_{\mu \, \nu} = \frac{1}{2}\epsilon_{\mu \, \nu \, \sigma \, \rho} G^{\sigma \, \rho}$, and for this model
\bea
c_{\tilde{G}}^\phi = - \frac{N \, g_3^2}{32 \, \sqrt{6} \, \pi^2}, \quad \, \quad c_{\tilde{B}}^\phi = - \frac{9 \, (a^2-b^2) \, N \, g_1^2}{80 \, \sqrt{6} \, \pi^2},
\eea
where $a,b$ are the hypercharges of the constituent particles.
In this case the decay to diphotons is considered as analogous to that of the neutral pion, where the decay is calculable and due to the chiral anomaly. 
Accompanying new scalar mesons of the new confining interactions are generically expected. These scalars will be bounded by the mixing constraints determined in the previous sections. 
The QCD example is the very wide $\sigma$ meson, which decays 
dominantly as $\sigma \rightarrow \pi \, \pi$, but does have a known decay into $\gamma \, \gamma$. We can develop a very rough understanding of the relationship between the couplings of a composite pseudoscalar to photons and those of a corresponding scalar on the basis of the constituent dynamics; one expects that the corresponding couplings are related by 
\bea
\frac{c_{\tilde{B}}^S}{c_{\tilde{B}}^\phi} \sim \frac{a^2+b^2}{a^2-b^2} \, \frac{\Psi^s(0)}{\Psi^\phi(0)}
\eea
\begin{figure}
\hspace{5cm}
\begin{tikzpicture}
\draw  [dashed] [thick] (-1.7,0) -- (-0.75,0);

\draw (180:0.75) -- +(30:1.75) ;
\draw  [->][ultra thick]  (70:0.55)  -- + (30:0.01) ;
 
 \draw (180:0.75) -- +(330:1.75) ;
\draw  [->][ultra thick]  (290:0.55)  -- + (330:-0.01) ;

\draw (50:1.2) -- +(270:1.75) ;
\draw  [->][ultra thick]  (0:0.8)  -- + (270:0.01) ;

\draw[decorate,decoration=snake] (50:1.2) -- +(0:1.1) ;
\draw[decorate,decoration=snake] (310:1.2) -- +(0:1.1) ;

\node [left][ultra thick] at (-1.7,0) {$\phi,S$};
\node [right][ultra thick] at (1.8,0.9) {$g,\gamma$};
\node [right][ultra thick] at (1.8,-0.95) {$g,\gamma$};
\node [right][ultra thick] at (-1.2,-0.5) {$\gamma^5, I$};
\end{tikzpicture}
\caption{Anomaly diagrams for $\phi$ and $s$.}
\end{figure}
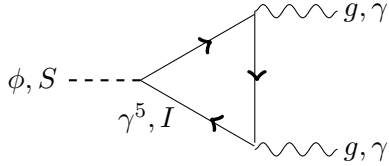
here $\Psi^s(0)$ and $\Psi^\phi(0)$ are the wavefunctions at the origin of the bound states. This is expected if the constituents mediating the coupling to two photons are identical, 
leading to the same loop function for two mesons.  This corresponds to considering a scalar with identical flavor quantum numbers as the pseudoscalar. In the pseudoscalar case there is an insertion of $\gamma^5$ in the diagram leading to the difference between squared hypercharges, while in the scalar case a unit matrix sums the squared charges, see Fig.4.

There is no reason to expect $\Psi^s(0) \ll \Psi^\phi(0)$ in general. The same reasoning applies to decays to $gg$. Mixing bounds then apply to the new scalar. Although $m_s$ can exceed $m_\phi$, the typical separation expected is $m_s/m_\phi \lesssim 4 \, \pi$.

\section{Conclusions}
We have examined the consistency of {run I} Higgs data and a putative diphoton excess at 750 {\rm GeV}, considering scalar and pseudoscalar states
that have an impact on lower energy phenomenology using the SMEFT formalism. We find that large mixings of a 750 {\rm GeV} state ({\it i.e. Wilson coefficients of the relevant operator $S H^\dagger \, H$ proximate to the cutoff scale}) are challenged by
these concerns, and have examined the corresponding naturalness bounds on the radiatively generated Wilson coefficient, due to the interactions required 
to produce the excess in diphotons. In general, we find that once a loop suppression of this Wilson coefficient is introduced, scalar models can be viable, and
pseudoscalar models are more protected from dangerous low-energy effects. One-loop matchings due to the pseudoscalar interactions do generate the operator
$O_{HB}= H^\dagger \, H \, B^{\mu \, \nu} \, B_{\mu \, \nu}$. The diphoton excess is not strongly challenged by consistency with lower energy data we have considered, in the simple scenarios we have examined.

\section*{Acknowledgements}

JC is grateful to the NBIA for its generous
hospitality during this work, which is also supported
by NSERC (Canada). MT and WS acknowledge generous support from the Villum Fonden. We thank Angelo Monteaux for a helpful comment,
and Zuowei Liu for useful discussions.

Note added: As well as the papers cited in the text, and utilized in this work, the following papers appeared on the 
archive discussing the $750 \, {\rm GeV}$ excess prior to this paper \cite{Alves:2015jgx,Bai:2015nbs,Agrawal:2015dbf,Ahmed:2015uqt,Chakrabortty:2015hff,Bian:2015kjt,Curtin:2015jcv,Chao:2015ttq,Demidov:2015zqn,No:2015bsn,Becirevic:2015fmu,Cox:2015ckc,Martinez:2015kmn,Kobakhidze:2015ldh,Matsuzaki:2015che,Cao:2015pto,Dutta:2015wqh,Petersson:2015mkr,Molinaro:2015cwg,Gupta:2015zzs,Bellazzini:2015nxw,Low:2015qep,Ellis:2015oso,McDermott:2015sck,Higaki:2015jag,DiChiara:2015vdm,Pilaftsis:2015ycr,Buttazzo:2015txu,Knapen:2015dap,Nakai:2015ptz,Angelescu:2015uiz,Backovic:2015fnp,Mambrini:2015wyu}.

\appendix
\section{One-loop Results}
Fig.2a gives the one-loop contribution to the Wilson coefficient $C_{HB}$ matching condition
\bea
\frac{ \Delta^a \, C_{HB}(m_s) \, g_1^2}{\Lambda^2}=  \frac{\lambda_3}{4 \, \pi^2} \left[\frac{5}{2} - \frac{\pi}{\sqrt{3}} \right] \, \frac{g_1^4 \, c_B^2}{\Lambda_\gamma^2},
\eea
while Fig.2b gives the contribution 
\bea
\frac{ \Delta^a \, C_{HB}(m_s) \, g_1^2}{\Lambda^2}= \frac{1}{4 \, \pi^2} \left[-\frac{5}{2} + \frac{\pi^2}{12} \right]\, \frac{g_1^6 \, y_h^2 \, c_B^2}{\Lambda_\gamma^2},
\eea
when calculating the unbroken phase of $\rm SU(2)_L \times U(1)_Y$ to simplify the matching. Note we take the real part of the amplitude in the matching as the Wilson coefficient of the Hermitian operators are real. Fig.2b vanishes for $C_{HG}$ while Fig.1a is the obvious modification of the quoted result for this operator.

\section*{B\quad The total photoproduction cross section}

Recent estimates of the combined inelastic-inelastic, elastic-inelastic and elastic-elastic photoproduction \cite{Fichet:2015vvy,Fichet:2016pvq,Csaki:2016raa} give a corrected $C_{\gamma\gamma}=78.3$, which is significantly higher than the estimate reported in \cite{Csaki:2015vek} (and initially used in this paper) for only the elastic production mechanism. Here we give the formulae arising from this total photoproduction. Utilizing this new coefficient, Equation \ref{dsigma2} now reads

\begin{equation}
\frac{\Delta \sigma(pp \rightarrow S \rightarrow \gamma \, \gamma)}{8 [fb]}\left(\frac{\Gamma_s}{45 \, {\rm GeV}} \right) \cong 6546 \, \left(\frac{m_s^2 \, c_B^2}{\Lambda_\gamma^2}\right)  \, \left[\left(\frac{m_s^2 \, c_G^2}{\Lambda_g^2} \right) +  3.5 \times 10^{-5} \, \left(\frac{m_s^2 \, c_B^2}{\Lambda_\gamma^2} \right)  \right],
\end{equation}
and Eqn. \ref{main:constraint}-\ref{widthfix} are $\mathcal{O}  \simeq 0.0005  \, \left(\kappa_\gamma -1 \right)^2 \left[\left(\kappa_g -1 \right)^2 + 6.2 \times 10^{-4} \, \left(\kappa_\gamma -1 \right)^2 \right]$
\begin{eqnarray}
\frac{\Delta\sigma\left(pp\to S\to\gamma\gamma\right)}{8[fb]}\left(\frac{\lambda\Lambda_c}{Nm_s}\right)^2&=& 0.06 \, N^2 \, \left(\kappa_\gamma-1\right)^2\frac{\left(\kappa_g-1\right)^2+6.2\times10^{-4}\left(\kappa_\gamma-1\right)^2}{\left(\kappa_g-1\right)^2+0.017\left(\kappa_\gamma-1\right)^2}, \nonumber \\
\frac{\Delta \sigma(pp \rightarrow S \rightarrow \gamma \, \gamma)}{8 [fb]}\left(\frac{\Gamma_s}{45 \, {\rm GeV}} \right)&=&\frac{1.7\times10^5}{\lambda_3^2}\left(\kappa_\gamma-1\right)\left(\left(\kappa_g-1\right)+2.2\times10^{-2}\left(\kappa_\gamma-1\right)\right). \nonumber \\
\end{eqnarray}
The resulting modified bounds are illustrated in Fig.6.
\begin{figure}[t]
  \centering
  \includegraphics[width=3.0in,height=3in]{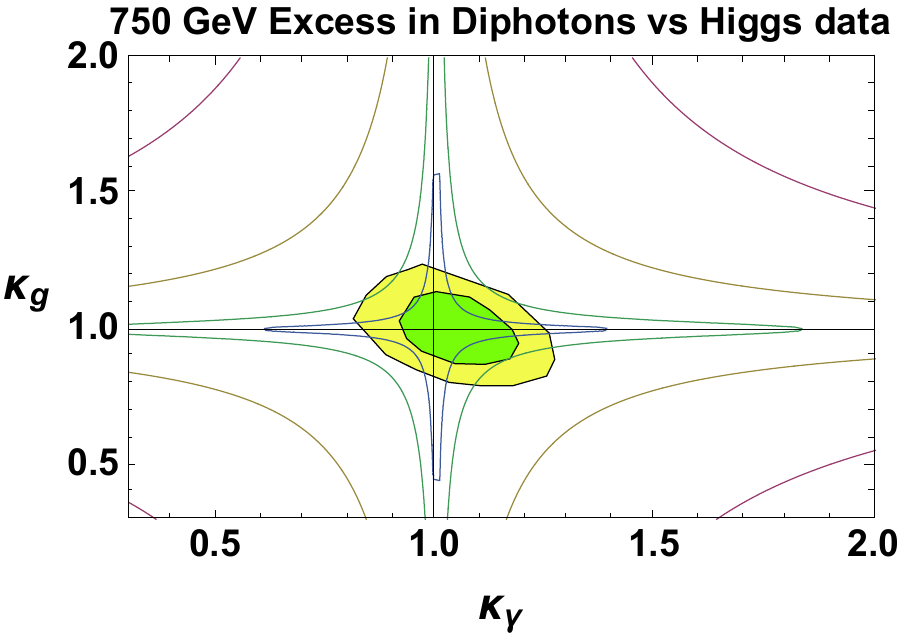}
    \includegraphics[width=3.0in,height=3in]{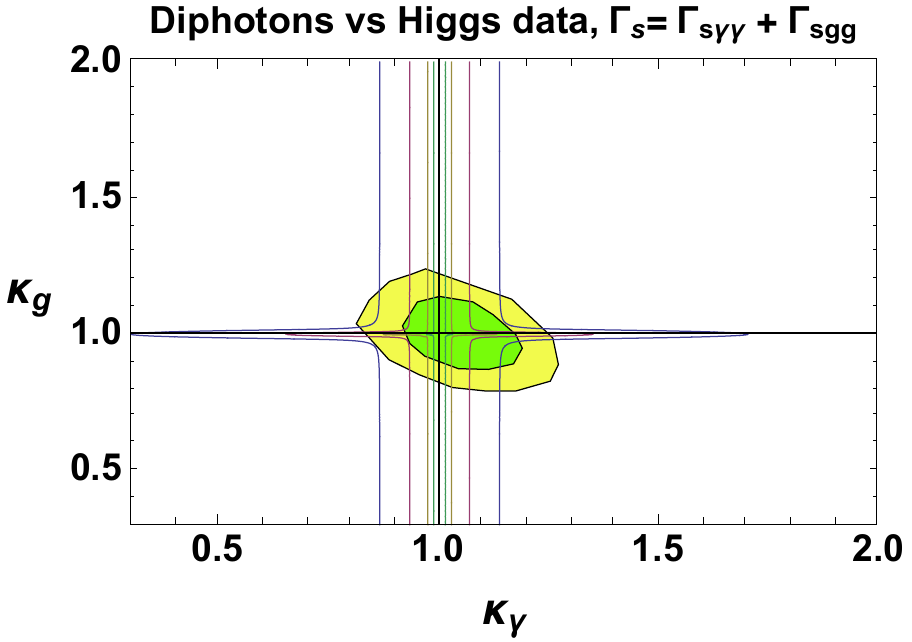}
  \caption{\label{Fig:nsvsr}Same as Figure 3, except for the total photoproduction cross section number in the Appendix. The curves in the left hand plot are again values of $\lambda \Lambda_c/N \times 750 \, {\rm GeV}  = \{0.1,0.05,0.02,0.01\}$ 
  coming in from the outermost curve in Eqn.\ref{main:constraint}. The right hand plot shows $\lambda \Lambda_c/N \times 750 \, {\rm GeV}   = \{0.1,0.05,0.02,0.01\}$. In both figures, the curves are overlaid on the
  $68\%$ and $95\%$ CL curves from the {run I} Atlas-CMS Higgs combination, fitting only to $\kappa_g, \kappa_\gamma$ \cite{ATLAS-CONF-2015-044}.}
\end{figure}

\bibliographystyle{JHEP}
\bibliography{RG}

\end{document}